%% file: Main.tex
\documentclass[prl,twocolumn,english,amsmath,amssymb,floatfix,superscriptaddress,longbibliography,preprintnumbers]{revtex4-2}

\usepackage{graphicx}
\usepackage{dcolumn}
\usepackage{bm}

\usepackage[utf8]{inputenc}
\usepackage[T1]{fontenc}
\usepackage{mathptmx}
\usepackage{etoolbox}
\usepackage{color}
\usepackage[colorlinks=true,citecolor=blue,urlcolor=blue]{hyperref}

\newcommand{\Vpol}{\mathrel{\updownarrow}}
\newcommand{\Hpol}{\mathrel{\leftrightarrow}}
\newcommand{\Dpol}{\mathrel{\text{$\nearrow$\llap{$\swarrow$}}}}
\newcommand{\Apol}{\mathrel{\text{$\nwarrow$\llap{$\searrow$}}}}

\begin{document}
\title{Direct Observation of Exceptional Points in Photonic Crystal by Cross-Polarization Imaging in Momentum Space}

\author{Viet Anh Nguyen}
\altaffiliation{The authors are equally contributed.}
\affiliation{College of Engineering and Computer Science, VinUniversity, Gia Lam district, Hanoi 14000, Vietnam}

\author{Viet Hoang Le}
\altaffiliation{The authors are equally contributed.}
\affiliation{College of Engineering and Computer Science, VinUniversity, Gia Lam district, Hanoi 14000, Vietnam}

\author{Lo\"{i}c Malgrey}
\affiliation{Univ Lyon, Ecole Centrale de Lyon, CNRS, INSA Lyon, Université Claude Bernard Lyon 1, CPE Lyon, CNRS, INL, UMR5270, 69130 Ecully, France}
\affiliation{Département de Physique, Ecole Normale Supérieure de Lyon,  46 Allée d’Italie, F 69342, Cedex 07 Lyon, France}

\author{Eirini Sarelli}
\affiliation{Univ Lyon, Ecole Centrale de Lyon, CNRS, INSA Lyon, Université Claude Bernard Lyon 1, CPE Lyon, CNRS, INL, UMR5270, 69130 Ecully, France}

\author{Dang-Khue Luu}
\affiliation{College of Engineering and Computer Science, VinUniversity, Gia Lam district, Hanoi 14000, Vietnam}

\author{Ha Linh Chu}
\affiliation{College of Engineering and Computer Science, VinUniversity, Gia Lam district, Hanoi 14000, Vietnam}

\author{Truong Tuan Vu}
\affiliation{College of Engineering and Computer Science, VinUniversity, Gia Lam district, Hanoi 14000, Vietnam}

\author{Cong Quang Tong}
\affiliation{Institute of Materials Science, VAST, 18 Hoang Quoc Viet, Ha Noi 11300, Vietnam}

\author{Vu Dinh Lam}
\affiliation{Graduate University of Science and Technology, Vietnam Academy of Science and Technology, 18 Hoang Quoc Viet, Hanoi, 11300, Vietnam}

\author{Christian Seassal}
\affiliation{Univ Lyon, Ecole Centrale de Lyon, CNRS, INSA Lyon, Université Claude Bernard Lyon 1, CPE Lyon, CNRS, INL, UMR5270, 69130 Ecully, France}

\author{Quynh Le-Van}
\email{quynh.lv@vinuni.edu.vn}
\affiliation{College of Engineering and Computer Science, VinUniversity, Gia Lam district, Hanoi 14000, Vietnam}

\author{Hai-Son Nguyen}
\email{hai-son.nguyen@ec-lyon.fr.}
\affiliation{Univ Lyon, Ecole Centrale de Lyon, CNRS, INSA Lyon, Université Claude Bernard Lyon 1, CPE Lyon, CNRS, INL, UMR5270, 69130 Ecully, France}
\affiliation{Institut Universitaire de France  (IUF), Paris, France}

\date{\today}

\begin{abstract}
This study explores exceptional points (EPs) in photonic crystals (PhCs) and introduces a novel method for their single-shot observation. Exceptional points are spectral singularities found in non-Hermitian systems, such as leaky PhC slabs. However, directly observing EPs in PhC systems using regular reflectivity spectroscopy is a considerable challenge due to interference between guided resonances and background signals. In this work, we present a simple, nondestructive technique that employs crossed polarizations to directly observe EPs in momentum-resolved resonant scattering. This approach effectively suppresses the background signal, enabling exclusive probing of the guided resonances where EPs manifest. Our results demonstrate the formation of EPs in both energy-momentum mapping and isofrequency imaging. All experimental findings align seamlessly with numerical simulations and analytical models. Our approach holds great potential as a robust tool for studying non-Hermitian physics in PhC platform.
\end{abstract}

\maketitle
Over the past decade, there have been remarkable developments in non-Hermitian physics within photonic platforms, encompassing both purely lossy systems and parity-time symmetric systems that exhibit a balance between optical gain and losses\cite{El-Ganainy2019,Yan2023,Nasari2023,Li2023}. These advancements have been possible due to the expertise in crafting synthetic optical materials at the subwavelength scale, which allows for precise modulation/tailoring of their gain and loss properties. The physics of these open systems can be depicted using effective non-Hermitian Hamiltonians\cite{rotter_non-hermitian_2009}, characterized by complex eigenvalues and non-orthogonal eigenvectors. Intriguingly, the degeneracies in a non-Hermitian Hamiltonian are termed Exceptional Points (EPs), where both the eigenvectors and eigenvalues converge. EPs represent singularities in non-Hermitian topology\cite{okuma2023,ding_non-hermitian_2022,Nasari2022,Li2023}, and their unique attributes have been instrumental in illustrating numerous non-trivial physical phenomena, including coupled waveguides \cite{benisty_transverse_2015,Goldzak2018,Khurgin_21,Schumer2022},  unidirectional reflectionless light propagation\cite{Lin2011,Regensburger2012}, loss-gain microresonators for lasers \cite{Peng2014,peng_loss-induced_2014,brandstetter_reversing_2014, feng_single-mode_2014,hodaei_parity-timesymmetric_2014,Shahmohammadi2022}, improved sensing sensitivity \cite{chen_exceptional_2017, hodaei_enhanced_2017, park_symmetry-breaking-induced_2020}, enhanced local density of optical states and spontaneous emission \cite{Pick2017,ferrier_unveiling_2022}.

Leaky resonances in Photonic Crystals (PhCs) serve as a versatile platform for studying non-Hermitian band structures and EP physics\cite{Fan2002,Liang2011,ZHOU20141}. By engineering energy-momentum dispersion and radiative losses in PhC systems, it becomes possible to generate EPs even in the absence of gain medium and material losses\cite{zhen_spawning_2015, Kaminski2017,Zhou2018,Bykov2018,lee_band_2019,lu_engineering_2020,Wu2020,ferrier_unveiling_2022}. The emergence of EPs within passive photonic crystals has paved new paths for investigating the captivating physics concealed within these structures. These include EP rings\cite{zhen_spawning_2015,Kaminski2017}, bulk Fermi arc\cite{zhen_spawning_2015}, non-Hermitian band inversion\cite{lee_band_2019,lu_engineering_2020,ferrier_unveiling_2022}, non-Hermitian coupling between orthogonal polarization modes\cite{Bykov2018,Wu2020}, and non-trivial propagation decay\cite{Wu2020}. However, demonstrating EPs in such PhCs is not straightforward, as shown by the detailed analysis required in angle-resolved reflectivity measurements\cite{zhen_spawning_2015}. This complexity stems from the Fano-shaped profile in  the reflectivity spectra, which arises from the coupling between guided resonances primarily confined within the PhC slab and background optical signals emanating from the substrate and the interfaces between layers. As a result, the presence of EPs tends to be obscured in angle-resolved reflectivity spectra, necessitating advanced analytical techniques. As a potential solution, coupling the EPs to a luminescent layer allows the detection of EPs through an enhancement of fluorescence in the energy-momentum spectra\cite{ferrier_unveiling_2022}. While this approach facilitates direct observation of EPs, the inclusion of an active material modifies the optical environment. It also introduces nonradiative losses, thereby significantly altering the properties of EPs from the pristine sample. Another tactic involves using a dielectric or metallic mirror as the sample substrate\cite{lu_engineering_2020,ferrier_unveiling_2022}. This leads to a Lorentz-shaped profile, as opposed to Fano ones, enabling direct observation of EPs. Nevertheless, requiring a mirror substrate imposes significant constraints and limits the design flexibility of the PhC. Therefore, it is imperative to develop a method for the direct observation of EPs in passive PhCs that maintains their inherent characteristics and imposes no restrictions on the stack design.

In this letter, we demonstrate that EPs in a passive PhC structure can be directly observed in angle-resolved reflectivity using a crossed-polarization setup. This technique, termed resonant scattering, effectively suppresses the specular reflection, thereby eliminating Fano interference and allowing for the exclusive probing of the PhC's resonant modes. EPs are discerned in single-shot imaging without any post-processing in both energy-momentum dispersion measurements and isofrequency contour measurements. The experimental findings align well with the results from numerical simulations. Our work introduces a straightforward and non-destructive method to monitor EPs in momentum space, further broadening the potential for resonance-enhanced scattering in photonic systems.

\begin{figure}[ht]
\centering
\includegraphics[width=1\linewidth]{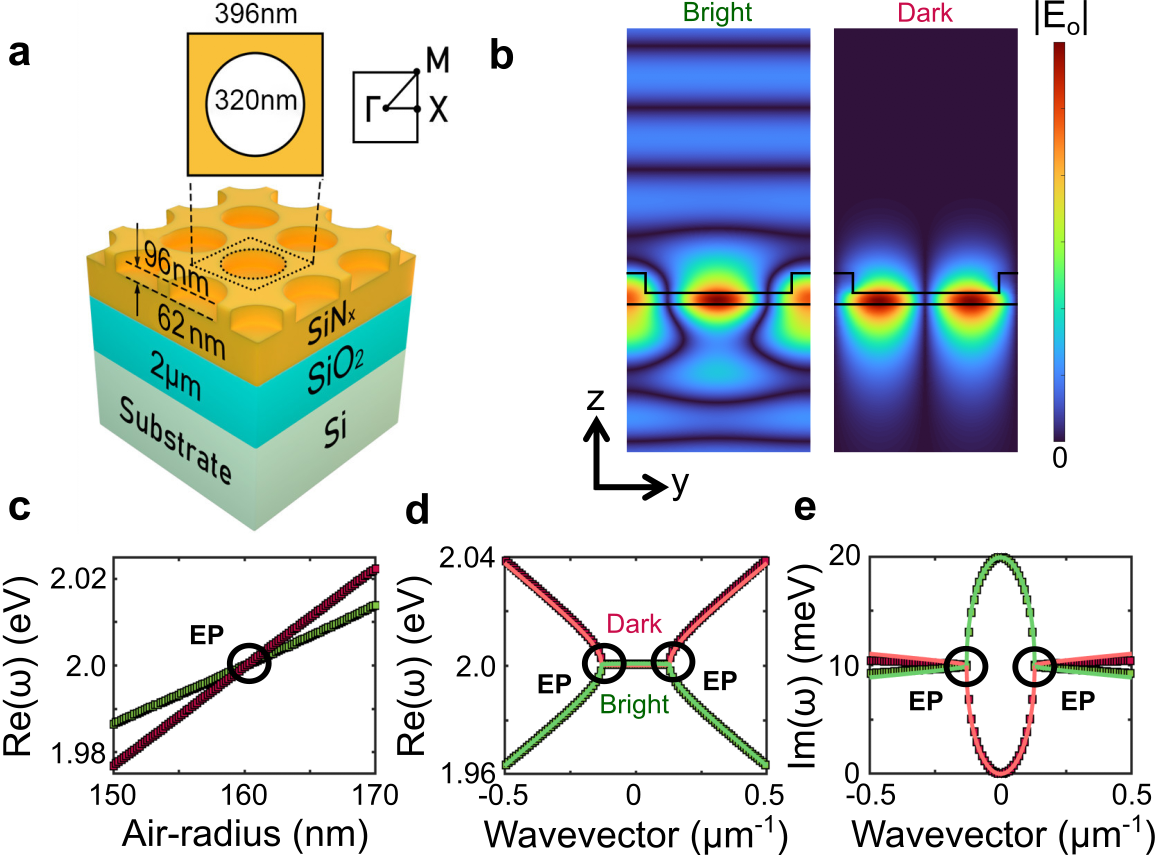}
\caption{\textbf{Sample design and photonic band structure}. a) Sketch of the sample design. The inset shows the Brillouin zone of the reciprocal lattice. b) The field intensity of the bright and dark modes at $k = 0$. c) The energy of bright and dark modes at $k$ = 0 for different hole radius. d) Real and e) imaginary parts of the energy eigenvalues along $\Gamma$M direction for structure with $r=r_{EP}=160$ nm. The green color corresponds to the bright mode while the red represents the dark modes. Analytical calculations are shown in solid line (pale red and pale green) and the numerical simulations are presented with square symbols (dark red and dark green).  The refractive indices of ${\text{SiN}}_{\text{x}}$ and ${\text{SiO}}_{\text{2}}$ used in the FEM simulations are 2.02 and 1.46, respectively. Analytical fits are obtained with $E_0=2.0$ eV, $\delta$ = 0.59 meV, $\gamma$ = 19.9 meV, $v=76.6$ meV.µm and $\sigma=-35$ meV.µm$^{-2}$. }
\label{fig1}
\end{figure}

The sample design, depicted in Fig.\ref{fig1}a, comprises a 158 nm thick SiN$_\text{x}$ layer, partially etched to form a square lattice of holes (96 nm depth and radius $r=160$ nm) with a period of 396 nm. This PhC slab is positioned on a silicon substrate with a 2µm-thick layer of SiO$_2$ serving as an optical spacer. In passive PhC slabs, EPs arise when two resonances of differing radiative losses coalesce\cite{zhen_spawning_2015,Bykov2018,ferrier_unveiling_2022}. In our design, these resonances encompass a leaky mode, termed a bright mode, and a non-radiating mode, named the dark mode\cite{ferrier_unveiling_2022,zhen_spawning_2015,Kaminski2017}. The bright (dark) mode is permitted (forbidden) to radiate into the far field due to its symmetric (anti-symmetric) near-field pattern. Figure~\ref{fig1}b displays the electric field distribution of these modes, derived from Finite Element Method (FEM) simulations using Comsol Multiphysics. It reveals that the electric field of the bright mode radiates into free space in the far field, whereas the field distribution of the dark mode is predominantly confined to the ${\text{SiN}}_{\text{x}}$ layer, not radiating into the far field. Notably, these modes resemble those previously documented for engineering exceptional rings\cite{zhen_spawning_2015,Kaminski2017}. At an oblique angle (i.e., nonzero momentum $k$), mirror symmetry is disrupted, causing the modes to hybridize. The effective Hamiltonian of the band structure near the $\Gamma$ point is given by\cite{ferrier_unveiling_2022}:
\begin{equation}\label{eq:Hamiltonian}
H(k)= 
\begin{pmatrix}
    {E_0 + \frac{\delta+\sigma k^{2}}{2}} & {v.k}  \\
    {v.k} & {E_0 -\frac{\delta+\sigma k^{2}}{2}}
\end{pmatrix}
+\begin{pmatrix}
	i\gamma & 0  \\
	0 & 0
\end{pmatrix}
\end{equation}
 In the Hermitian term (i.e. first term) of \eqref{eq:Hamiltonian},  $E_0$ is the offset energy of the bands, $\delta$ is the frequency detuning between the bright mode and the dark mode at $k=0$; and $v, \sigma$ are coefficients of second-order $k.p$ perturbation theory. In the non-Hermitian term (i.e. second term) of \eqref{eq:Hamiltonian}, $\gamma$ corresponds to the radiative losses of the bright mode, given by: 
 \begin{equation}\label{eq:eigenvalues}
 E_{\pm}(k)=E_{0}+i\frac{\gamma}{2} \pm \sqrt{\left(\frac{\delta+\sigma k^{2}+i\gamma}{2}\right)^{2}+v^{2} k^{2}}
\end{equation}
  Interestingly, when the detuning $\delta$ is tuned to $\delta_{EP}=-\frac{\sigma\gamma^{2}}{4v^2}\approx 0$, the two eigenvalues coalesce to $E_{EP}=E_{0}+i\frac{\gamma}{2}$ at $k_{EP}=\pm\frac{\gamma}{2v}$ and EPs are formed. This configuration can be achieved in our design since $\delta$ can be finely tuned  by sweeping the air-radius of the hole lattice. Indeed, Figure~\ref{fig1}c presents FEM results of the energy of the two modes at $k=0$ as functions of the air radius $r$. It shows that the zero-detuning configuration is achieved for radius $r_{EP}=160$ nm. Figures~\ref{fig1}d,e presents the  real (Fig.~\ref{fig1}d) and imaginary (Fig.~\ref{fig1}e) parts of two complex eigenvalues obtained by FEM simulations along $\Gamma$M direction of the Brillouin zone when $r=r_{EP}$. These numerical results are perfectly fitted by the analytical model \eqref{eq:eigenvalues}. They clearly evidence the formation of EPs on  $\Gamma {\text{M}}$ path at $k_{EP} = \pm $0.13 µm $^{-1}$. The dark and bright modes can also be inferred in Fig.~\ref{fig1}e in which at $k=0$, the imaginary of the dark mode reduces to zero whereas the bright mode exhibits the highest loss. The calculated band structures in both directions $\Gamma $M and $\Gamma $X can be found in supplemental information (SI). 

\begin{figure}[ht!]
\includegraphics[width=1\linewidth]{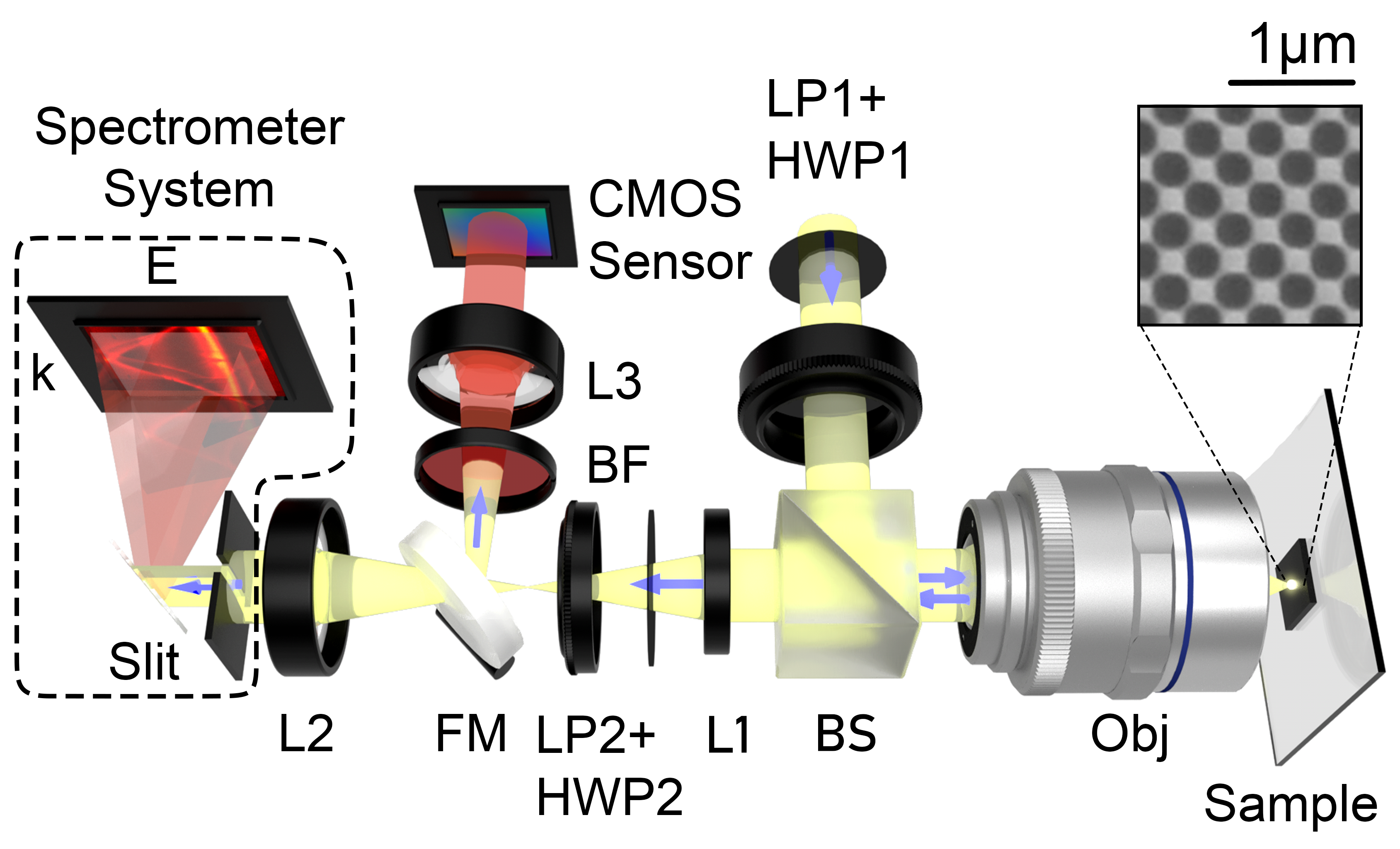}
\caption{\textbf{Experimental setup}. Illustration of the optical measurement system. OL: objective lens. BS: beamsplitter. LP: linear polarizer. HWP: half-wave plate. L1, L2, L3: lens system. FM: flippable mirror. BF: band-pass filter ($610\text{nm} \pm 10\text{nm}$). Inset:  SEM image of the sample's surface.}\label{fig2}
\end{figure}

The sample is fabricated using Plasma Enhanced Chemical Vapor Deposition to deposit SiN$_\text{x}$. This is followed by laser interference lithography and dry etching to pattern the SiN$_\text{x}$ layer into PhC structures. Further fabrication details are provided in Section I of the SI. A scanning electron microscopy (SEM) image of the final sample is presented in the inset of Fig.\ref{fig2}.

To solely probe the guided resonances in our PhC structure, we employ the resonant scattering measurement technique that involves reflectivity/transmission spectroscopy in  a cross-polarization setup\cite{McCutcheon2005,Galli2009,Regan2016,Nazirizadeh2008,Nazirizadeh2008bis,Lueder2020}. While no previous studies have utilized this method to unveil EPs, it has been effectively used to discern bare photonic modes for localized states in PhC cavities\cite{McCutcheon2005,Galli2009,Deotare2009,Ma2023} and the band structures of guided mode resonances\cite{Nazirizadeh2008,Bajoni2009,Dang2022}. The underlying principles of the resonant scattering technique, used to determine the spectral position of guided resonances in PhC, are discussed extensively in earlier reports\cite{Nazirizadeh2008,Nazirizadeh2008bis,Lueder2020}. In essence, the cross-polarization arrangement in resonant scattering spectroscopy results in the suppression of specular reflection/transmission and its interference with light scattered from the guided resonances. Consequently, the spectra exhibit a zero baseline and symmetric profile peaks, diverging from the Fano profile. This facilitates direct insight into the spectral position and line-width of the guided resonance.

The optical setup for the resonant scattering measurement in momentum space is illustrated in Fig.~\ref{fig2}. Broadband light (Thorlabs SLS202C) is focused onto the sample through a microscope objective (OptoSigma 50X, NA = 0.55). Polarization is controlled using a linear polarizer (LP1) and a half-wave plate (HWP1). Reflected light is collected by the same objective and separated by using a beam splitter (BS). A Fourier lens (L1) images the objective's back focal plane. For resonant scattering, background signals are filtered with another polarization set (LP2 and HWP2), while regular reflectivity omits this step. For momentum-resolved spectroscopy, light is focused via lens (L2) onto a monochromator (Princeton Instrument HRS-300) slit, then relayed to a 2D spectroscopic camera (Kuro model). This setup maps the energy-momentum band structure in one shot, with the direction determined by the monochromator slit. For isofrequency Fourier imaging, a band-pass filter (BF) sets the wavelength, and a flipped mirror (FM) projects the back-focal plane to a CMOS camera (Caltex). This maps photonic modes in momentum space ($k_x$, $k_y$) at specific energy with a single-shot measurement.

Four linear polarizations are defined: horizontal ($\Hpol$), vertical ($\Vpol$), diagonal ($\Dpol$), and antidiagonal ($\Apol$). These polarizations correspond to angles of 0$^\text{o}$, 90$^\text{o}$, 45$^\text{o}$, and -45$^\text{o}$, respectively, between the axes of the polarizers (LP1, LP2) and the x-axis of the sample. Resonant scattering experiments employ two configurations: the crossed polarizations of $\left[\Vpol,\Hpol\right]$ and $\left[\Apol,\Dpol\right]$. In the $\left[\Vpol,\Hpol\right]$ setup, light polarized in $\Vpol$ is incident on the sample, and the reflected light is analyzed in $\Hpol$ polarization. Conversely, in the $\left[\Apol,\Dpol\right]$ configuration, the sample is illuminated with $\Apol$-polarized light and the reflection is analyzed in $\Dpol$ polarization.

\begin{figure}[ht!]
\includegraphics[width=1\linewidth]{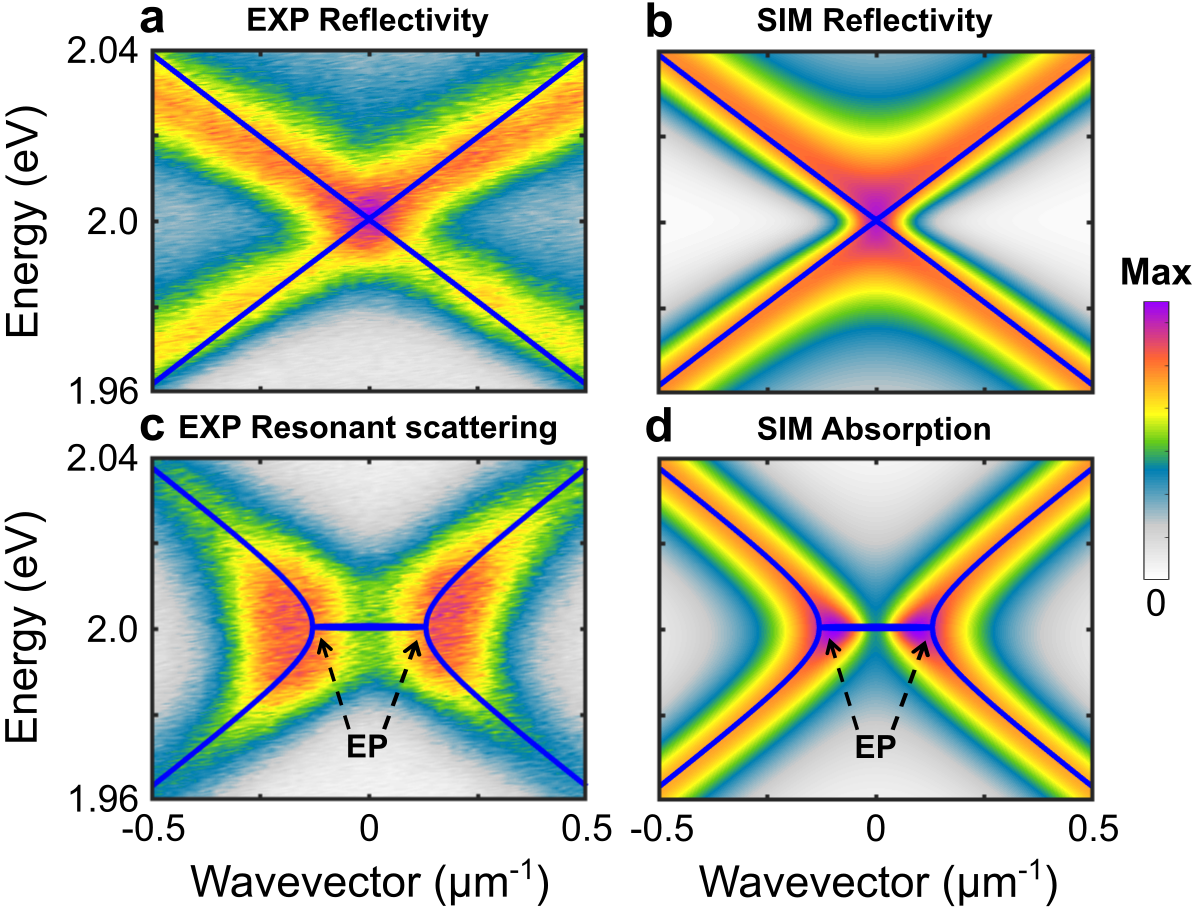}
\caption{\textbf{Angle-resolved reflectivity measurement}. a) Experimental and b) FEM simulation results of regular reflectivity with $\Dpol$-polarized excitation. c) Experimental results of resonant scattering measurement in the $\left[\Vpol,\Hpol\right]$ configuration. d) FEM simulation results of absorption with $\Dpol$-polarized excitation.  The blue solid line indicates analytical calculation of the band structure with $\gamma=0$ in (a,b), and $\gamma=19.9$ meV in (c,d). Each single shot of the measured spectrum is integrated for 100 ms and the monochromator slit is set at 100 µm. For the absorption simulation, an imaginary component of 0.005 has been added to the refractive index of SiN$_\text{x}$.}\label{fig3}
\end{figure}

We first perform a regular energy-momentum-resolved reflectivity experiment. The sample is oriented so that its x-axis is 45$^\text{o}$ tilted with respect to the slit of the monochromator, so the collection plane corresponds to the $\Gamma$M direction. Since the two modes of interest are of s-polarized light\cite{zhen_spawning_2015,Kaminski2017}, their polarization for this collection plane is purely $\Dpol$. Figure~\ref{fig3}a presents the experimental result of regular reflectivity for $\Dpol$ excitation. The corresponding numerical result by FEM is presented in Fig.\ref{fig3}b. Following the reflectivity peak, both experimental and numerical results show Dirac cone degeneracy at $k=0$ instead of branch degeneracy of EPs. Indeed, they are nicely fitted by the band structure of the Hermitian term of \eqref{eq:Hamiltonian}, given by $E_0 \pm v.k$ at zero-detuning and in the vicinity of $k=0$ (see solid blue curves in Figs.\ref{fig3}a,b). Therefore, regular reflectivity cannot provide a direct observation of non-Hermitian degeneracy.  These findings are corroborated by the studies of Zhen et al.\cite{zhen_spawning_2015}. In their earlier research, reflectivity mapping only shows Hermitian degeneracy (i.e. Dirac cones) and EPs only came to light after tedious analysis of reflectivity spectra and fitting with a theoretical model. 

Next, we undertake resonant scattering experiments in the $\left[\Vpol,\Hpol\right]$ configuration. This configuration is chosen since it enables the probing of our photonic modes of pure $\Dpol$ polarization (i.e., $\Dpol$ has a non-zero projection in both $\Vpol$ and $\Hpol$). As mentioned above, the crossed polarization setup suppresses specular signals and Fano features that hindered the EPs observation in the reflectivity measurements of Figs.\ref{fig3}a,b. Therefore, EPs can be directly revealed in the mapping of the resonant scattering experiment. Indeed, Figure\ref{fig3}c presents the experimental results of the energy-momentum-resolved resonant scattering measurement. The real part of eigenmodes from \eqref{eq:eigenvalues} is plotted over the experimental data. It shows that the non-Hermitian bands are revealed in this single-shot experiment by simply monitoring the intensity peaks. In particular, the branch points and momentum gap, hallmarks of EPs, are clearly evidenced, without using any post-treatment. We note that the dominant scattering signal at EPs (Figs.\ref{fig3}c) is due to the enhancement of local density of states at these spectral singularities\cite{Pick2017,ferrier_unveiling_2022}.These features are in agreement with the simulation results of absorption when the structure is excited with $\Dpol$-polarized light (Fig.\ref{fig3}d). It confirms that resonant scattering techniques can be used to probe EPs in PhC in a similar fashion as active measurements (i.e., photoluminescence or absorption). The broadening of experimental data (Figs.\ref{fig3}c) with respect to the FEM results (Figs.\ref{fig3}d) can be attributed to an inhomogeneous broadening effect, as discussed in the following.  

\begin{figure}
\includegraphics[width=1\linewidth]{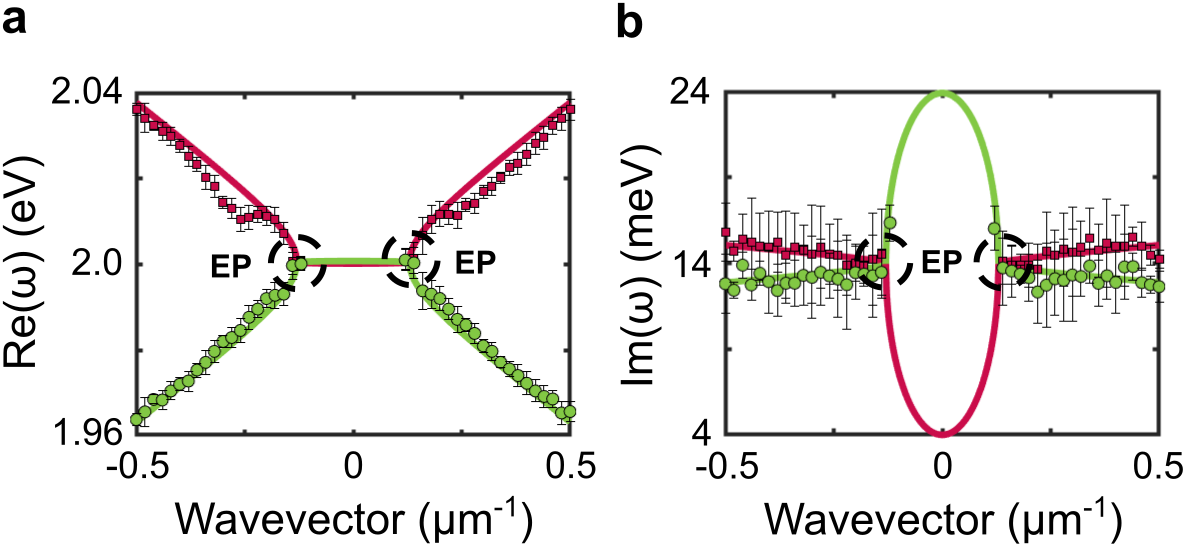}
\caption{\textbf{Quantitative analysis of EP formation}. a) Real and b) imaginary part of the eigenmodes. Symbols correspond to experimental data extracted from the fitting of resonant scattering spectra. Solid lines correspond to analytical calculation with inhomogenous broadening $\gamma_i=4$ meV included. The green color corresponds to the bright mode while the red represents the dark modes.}\label{fig4}
\end{figure}
To analyze quantitatively the formation of EPs in Fig.\ref{fig3}c, we extract the real and imaginary parts of the eigenvalues from resonant scattering measurements. This is achieved by fitting the resonances in the spectrum for each wavevector. It is important to note the presence of an inhomogeneous broadening in the measured spectra. This broadening is due to spatial variation of the hole radius, an undesired feature that is typical of laser interference lithography\cite{Capraro2023}. To account for this inhomogeneous broadening, we used Gaussian profiles instead of Lorentzian ones to fit the resonant scattering spectra. Examples of fitted spectra are shown in Fig. S4 of the SI. Figures~\ref{fig4}a and \ref{fig4}b show the real (determined by the position of the resonance) and imaginary (extracted from the linewidth of the resonance) parts of the eigenvalues of the band diagram, respectively. Importantly, a simultaneous degeneracy of real and imaginary parts is evidenced at $k$ = $\pm$0.13 µm$^{-1}$, thus confirming EPs formation at these wavevectors. We note that the extracted imaginary part, shown in Fig.\ref{fig4}b, exhibits an offset of 4 meV compared to that of the numerical design (Fig.\ref{fig1}e). This offset accounts for the inhomogeneous broadening discussed above. One may include this broadening into the analytical model by adding an offset $i\gamma_i$ to the diagonal coefficients of the non-Hermitian term in \eqref{eq:Hamiltonian}, with $\gamma_i=4$ meV. Once this offset is included, the experimental values align closely with the analytical model (see Figs.\ref{fig4}a,b). Further details of the analytical model, including inhomogeneous broadening, are discussed in Fig. S5 of the SI.

\begin{figure}
\includegraphics[width=1\linewidth]{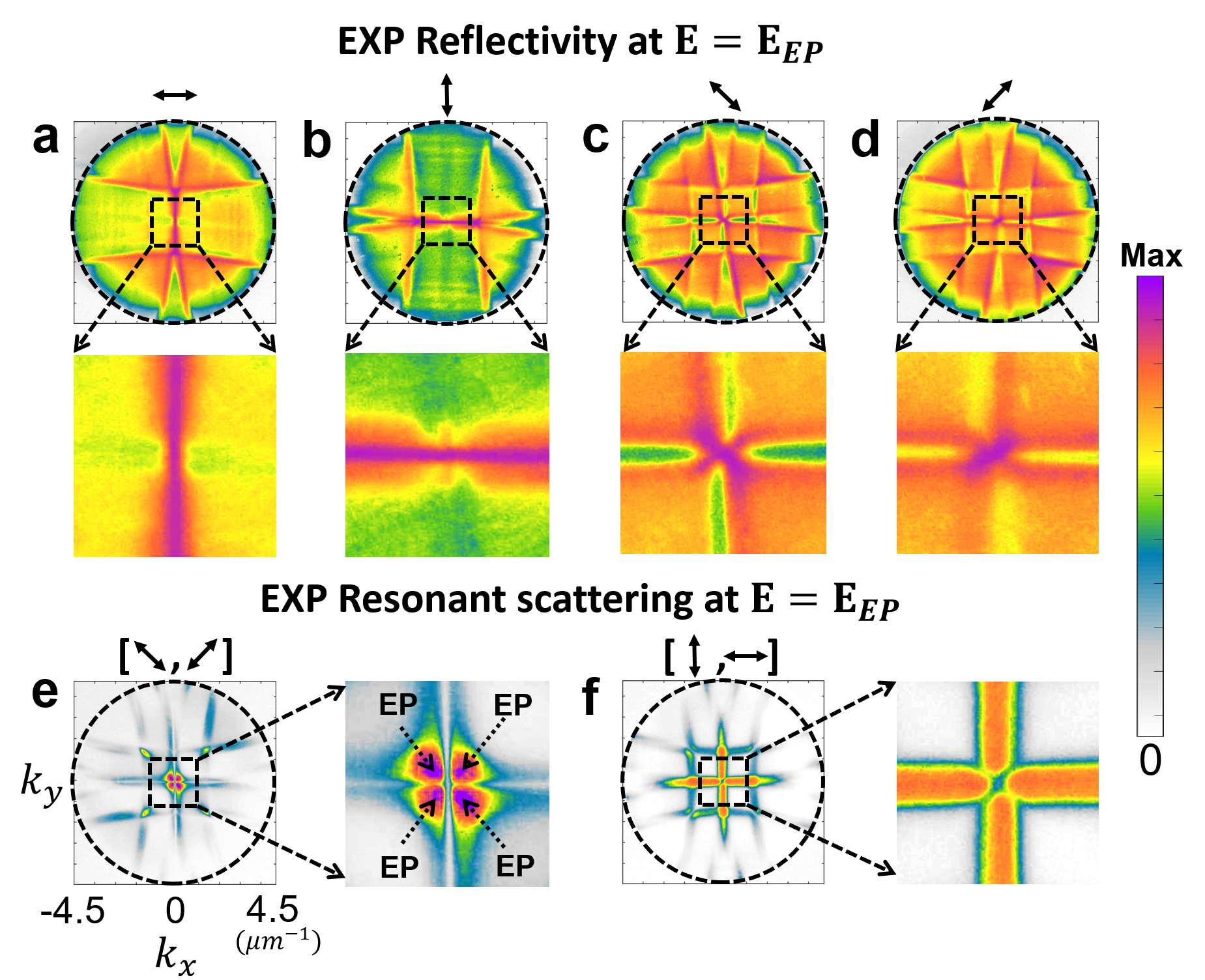}
\caption{\textbf{Fourier imaging at the EPs energy}. a-d) Experimental results of the regular reflectivity image for four different polarizations of excitation light. e,f) Experimental results of the resonant scattering image in two different polarization configurations.  The magnified views close to the $\Gamma$ point of corresponding Fourier images are shown right below (a-d) and on their right side (e,f).  The black dashed circle indicates the limit of the field of view, which corresponds to the numerical aperture NA = 0.55 of the microscope objective. The scale in $k_x$ is equivalent to $k_y$ for all the images. The EPs are points indicated by the black arrow in the magnified picture of (e).}\label{fig5}
\end{figure}

Finally, we perform Fourier imaging in the ($k_x$,$k_y$) plane at the EPs energy for different polarization configurations for regular reflectivity (Figs.\ref{fig5}a-d) and resonant scattering (Figs.\ref{fig5}e,f) experiments. This is achieved by adding a bandpass filter ($\lambda$=610 nm or $E=2.0$ eV) in the collection path to extract only spectral signals in the vicinity of the EPs and imaging the back focal plane in various polarization configurations. We now focus on the zoom-in images in the vicinity of the $\Gamma$ point (bottom panels of Figs.\ref{fig5}a-d, and right panels of Figs.\ref{fig5}e,f). These results clearly show that regular reflectivity mappings exhibit non-zero background from specular reflection, while the background is completely suppressed in resonant scattering mappings. Moreover, all isofrequency maxima curves in Figs.\ref{fig5}a-d go through the $\Gamma$  point, suggesting a degeneracy at $k=0$. This is consistent with the results from the energy-momentum measurements of Fig.\ref{fig3}a previously discussed: regular reflectivity only reveals the Hermitian version of the band structures. On the other hand, four local maxima corresponding to four EPs along $\Gamma$M  directions  ($\pm 45 ^\text{o}$ planes) are prominently observed in the resonant scattering mapping of the $[\Vpol,\Hpol]$ configuration (Fig.\ref{fig5}e). Such an observation is possible thanks to the local enhancement of the scattering signal at EPs. We also note that horizontal and vertical isofrequency segments are evident in the resonant scattering mapping of the $[\Apol,\Dpol]$ configuration (Fig.\ref{fig5}f). These segments correspond to the slow-light upper-band along the $\Gamma$X direction (more details of the band structure along $\Gamma$X are provided in Fig.S2 of the SI).

In conclusion, we have introduced a new approach for directly observing EPs in PhC structures of hole lattice. By leveraging the crossed polarization technique in momentum-resolved reflectivity, we effectively circumvented the challenges that predominantly stem from the complex interplay between guided resonances and background signals. Our experimental results of momentum-resolved resonant scattering and Fourier images validate the formation of EPs in a patterned SiN$_\text{x}$ slab, perfectly aligning with both theoretical predictions and simulations. This streamlined technique, adaptable across various PhC stacks and geometries, promises a powerful asset for diving deeper into non-Hermitian photonics and topology, particularly expanding the horizon of discoveries and applications in advanced photonic platforms.

\begin{acknowledgments}
The authors acknowledge the financial support of VinUniversity's seed grant under Grant No. VUNI.2021.SG10, VinUniversity-UIUC Smarthealth Center under Grant No. 8647, and Vingroup Innovation Foundation under Grant VINIF.2021.DA00169.
\end{acknowledgments}

\section*{Data Availability Statement}
The data that support the findings of this study are available from the corresponding author upon reasonable request.

\bibliography{EPs}

\include{SI}

\end{document}

%% file: SI.tex
\setcounter{equation}{0}
\setcounter{figure}{0}
\setcounter{table}{0}
\setcounter{page}{1}

\renewcommand{\theequation}{S\arabic{equation}}
\renewcommand{\thefigure}{S\arabic{figure}}
\renewcommand{\bibnumfmt}[1]{[S#1]}
\renewcommand{\vec}[1]{\boldsymbol{#1}}

\onecolumngrid 
 \begin{center}
\Large{--- SUPPLEMENTAL INFORMATION ---}
\end{center}
\normalsize{}
\section{Sample fabrication}
\begin{figure}[ht]
\centering
\includegraphics[width=0.4\linewidth]{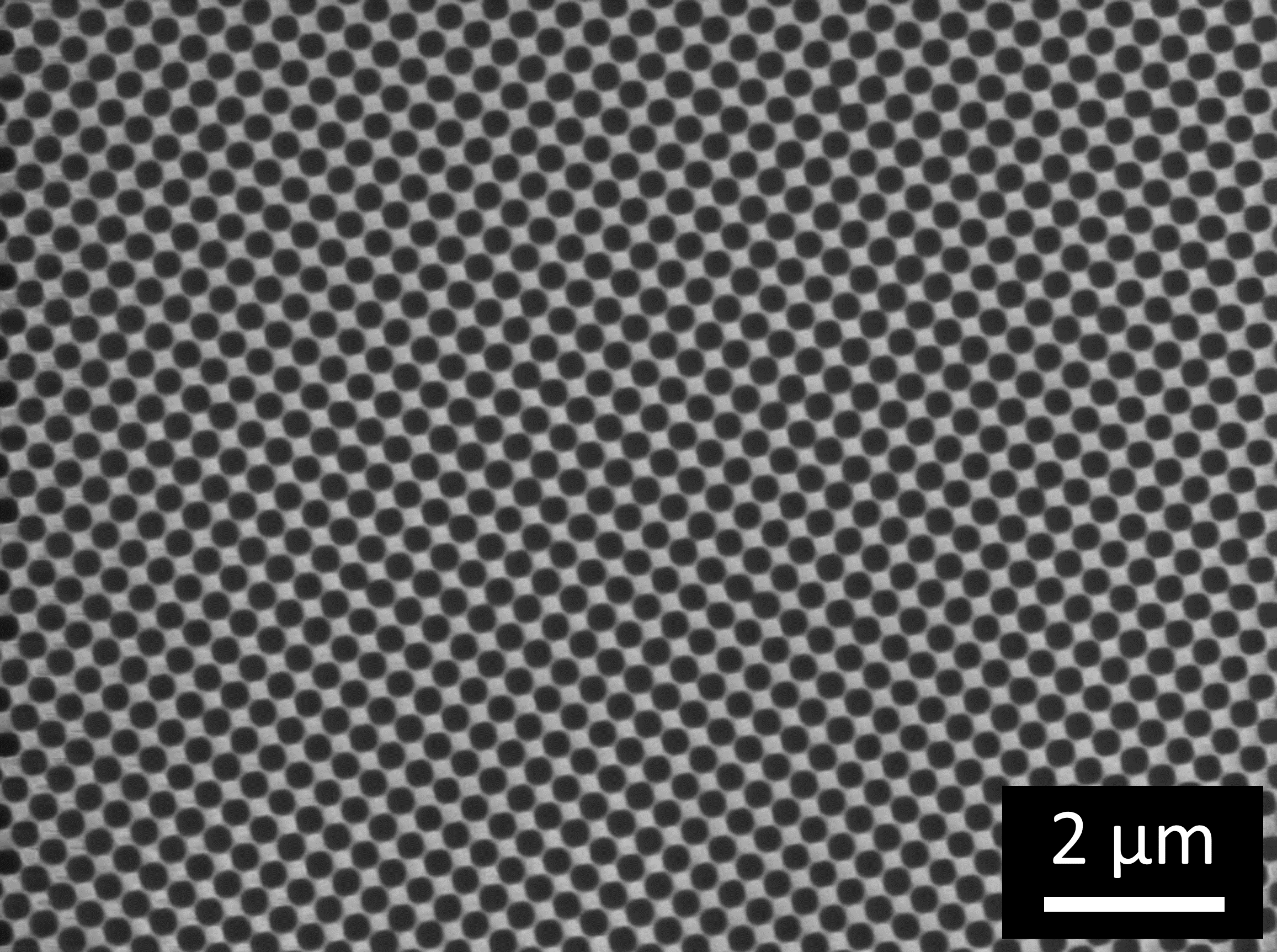}
\caption{\textbf{Scanning Electron Microscopy image of the fabricated sample}. A crop version of this image is shown in the inset of Fig.2 in the main text.}
\label{figS1}
\end{figure}

The designed structure is fabricated by depositing a 158 nm-thick layer of ${\text{SiN}}{\text{x}}$ directly onto a 2 $\mu \text{m}$-thick ${\text{SiO}}_{\text{2}}$ layer on a Si substrate using the Plasma Enhanced Chemical Vapor Deposition technique. We employ laser interference lithography to pattern the negative resist film (maN-2043). Post-development, the structure is transferred to ${\text{SiN}}{\text{x}}$ via Reactive Ion Etching using the Corial 200FA system, leveraging a gas mixture of $\text{CH}\text{4}$, $\text{SF}\text{6}$, and $\text{CHF}\text{3}$. This process lasts for nine minutes under a pressure of 100 mTorr at 20$^\text{o}$C. Subsequently, the partly etched ${\text{SiN}}_{\text{x}}$ final structure is attained by removing the resist using oxygen plasma. A scanning electron microscope (SEM) image of the fabricated sample is shown in Fig.~\ref{figS1}
\section{Photonic band structure}
\begin{figure}[ht]
\centering
\includegraphics[width=0.55\linewidth]{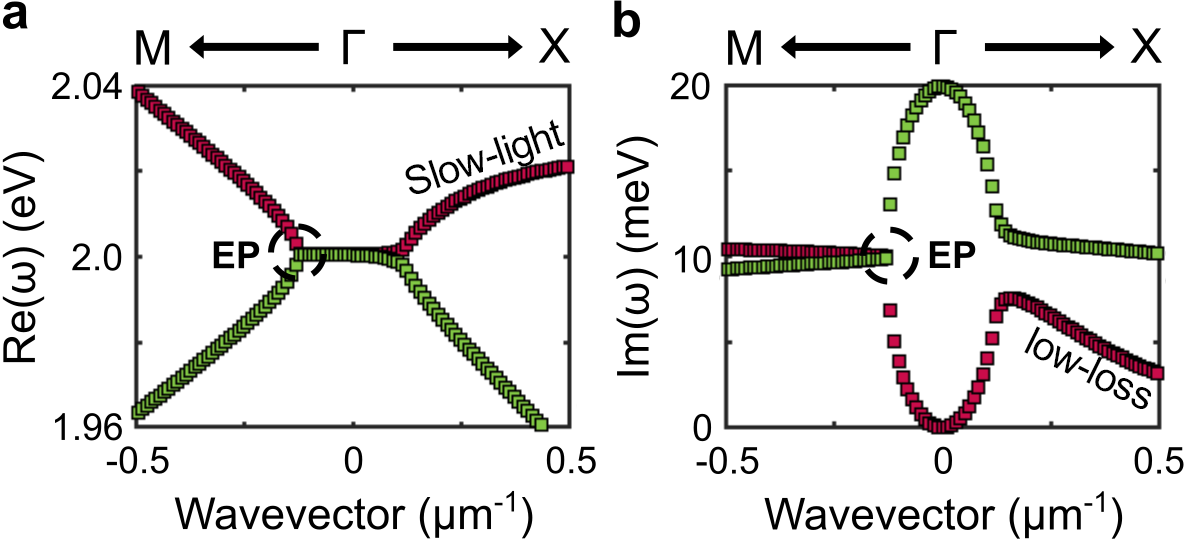}
\caption{\textbf{Band diagram of the structure}. a) and b) are real and imaginary parts of bright mode (green) and dark mode (red). The real and imaginary parts are coalesced in $\Gamma$M direction through a mark of open dash circle.}
\label{figS2}
\end{figure}

Figure~\ref{figS2} represents the results of the real (Fig.~\ref{figS2}a) and imaginary (Fig.~\ref{figS2}b) part of the photonic modes along both $\Gamma$X and $\Gamma$M directions. These results are obtained via FEM simulations for s-polarized modes (i.e. transverse electric modes, corresponding to $\Hpol$ for $\Gamma$X direction, and  $\Dpol$ for $\Gamma$M direction ). It shows that the EPs (degeneracy of both real and imaginary part of eigenvalues at the same point) occur only in $\Gamma$M  direction.  Interestingly, the band structure along $\Gamma$M is symmetric with respect to $E_{EP}$ for the real part and $\gamma_{EP}$ for the imaginary part. The energy-momentum dispersion along $\Gamma$M exhibits linear dispersion of opposite group velocities as soon as $|k|$>$|k_{EP}$. On the other hand, the band structure along $\Gamma$X does not possess the "particle-hole"-like symmetry as the one along $\Gamma$M  direction. In particular, the group velocity and the line-width of the upper-band decrease rapidly when moving out of the $\Gamma$ point along $\Gamma$X direction. Therefore, one may consider the upper-band as a slow-light mode with low-loss. The isofrequency segments along $\Gamma$X directions observed in Fig.5f of the main text correspond to this low-dispersive mode.

\begin{figure}[ht!]

\includegraphics[width=0.8\linewidth]{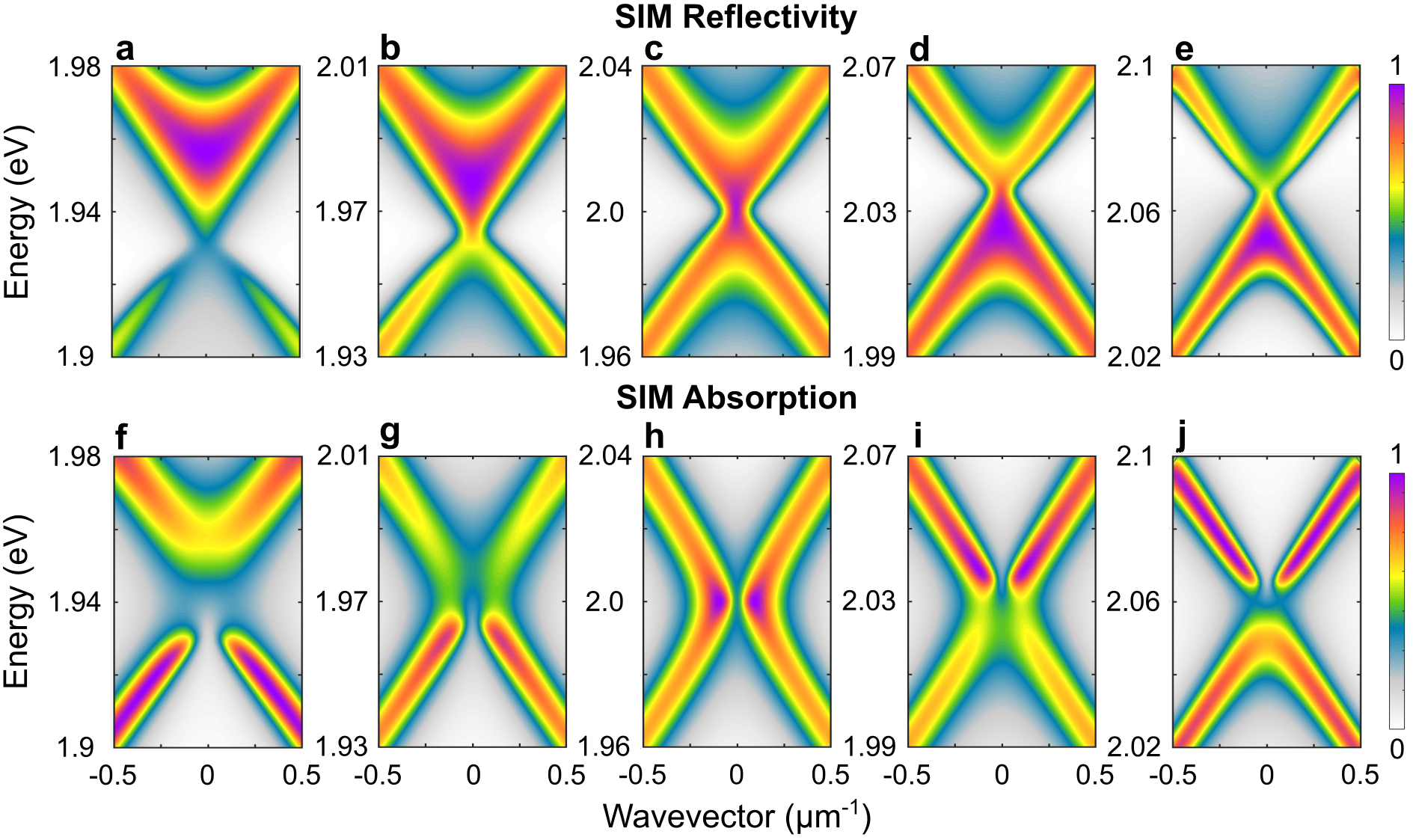}
\caption{\textbf{Angle-resolved reflectivity and absorption spectra for different hole radius}. a,f) $r = 0.8r_{\text{EP}}$, b,g) $r=0.9r_{\text{EP}}$,  c,h) $r = r_{\text{EP}}$,  d,i) $r = 1.1r_{\text{EP}}$ and e,j) $r = 1.2r_{\text{EP}}$.The upper panel shows regular reflectivity simulations and the lower panel shows angle-resolved absorption spectra simulations.} \label{figS3} 
\end{figure}

\section{Angle-resolved absorption simulation with different hole radius}
As discussed in the main text, the detuning between the dark and bright mode can be finely tuned by sweeping the hole radius $r$. EPs configuration corresponds to $r=r_{EP}=160$ nm. This effect is illustrated in Fig.~\ref{figS3}, showing energy-momentum mappings of the regular reflectivity (Fig.~\ref{figS3}a-e) and absorption (Fig.~\ref{figS3}f-j) along $\Gamma$M direction with $\Dpol$-polarized excitation for five different designs: $r=0.8r_{EP}$ (Fig.~\ref{figS3}a,f), $r=0.9r_{EP}$ (Fig.~\ref{figS3}b,g), $r=r_{EP}$ (Fig.~\ref{figS3}c,h), $r=1.1r_{EP}$ (Fig.~\ref{figS3}d,i), and $r=1.2r_{EP}$ (Fig.~\ref{figS3}e,j). These results are obtained by FEM simulations. The band inversion effect is clearly evidenced: the dark mode is in lower-band when $r<r_{EP}$, coalesces with the bright mode to make EPs when $r=<r_{EP}$, then is in the upper-band when $r>r_{EP}$. Interestingly, these results show that while the EPs cannot be revealed by regular reflectivity (Fig.~\ref{figS3}c) and can only be observed in absorption (Fig.~\ref{figS3}h), we can distinguish the dark and bright modes when the are not coalesce in both regular reflectivity and absorption spectra. 
\section{Gaussian fittings for experimental resonant scattering spectra}
\begin{figure}[ht!]
\includegraphics[width=0.8\linewidth]{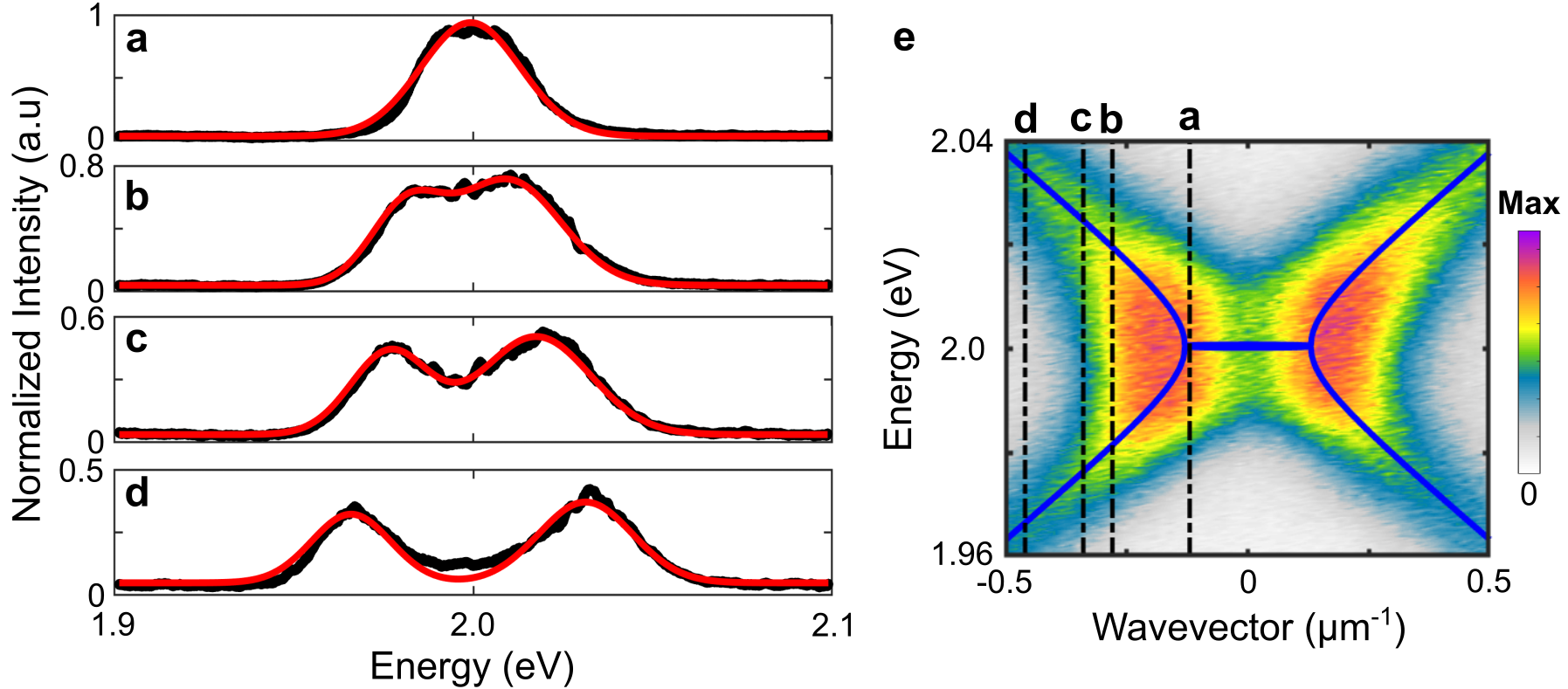}
\caption{\textbf{Gaussian Fitting of resonant scattering spectrum}. Data measurement (black) and Gaussian fitting (red) for a) $k =$ -0.12 µm$^{-1}$, b) $k =$ -0.28 µm$^{-1}$, c) $k =$ -0.34 µm$^{-1}$ and d) $k =$ -0.46 µm$^{-1}$. e) The position of these slices on the resonant scattering spectrum.}\label{figS4}
\end{figure}

Spatial variation of patterned features is a well-known undesired effect of laser inteferental lithography\cite{Capraro2023}. Such a slow variation leads to an inhomogenous broadening of photonic resonances. Thus Gaussian profiles are more adapted to fit the resonant scattering spectra instead of Lorentzian ones. Figure~\ref{figS4} present resonant-scattering spectra at three different wavectors. It shows a very good aggrement between the experimental data and Gaussian fittings.

\section{Inhomogeneous broadening of the photonic modes}
\begin{figure}[ht!]
\includegraphics[width=0.4\linewidth]{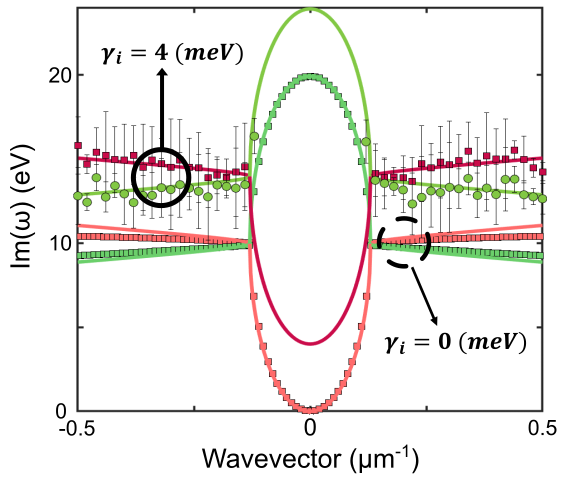}
\caption{\textbf{Inhomogeneous broadening of the photonic modes}. Comparison between imaginary parts obtained by analytical model, FEM simulation, and experimental results. Green and red bands are the results of the experimental results and analytical model with $\gamma_{i}$ = 4 meV. Pale red and pale green bands are FEM simulation results and analytical model with $\gamma_{i}$ = 0 meV. Solid lines correspond to analytical models and symbols correspond to experimental and simulation results.}\label{figS5}
\end{figure}

 The inhomogeneous broadening can be included in the non-Hermitian effective Hamiltonian by adding an offset loss $\gamma_i$ in the anti-Hermitian term:
\begin{equation}\label{eq:newHamiltonian}
H(k)= 
\begin{pmatrix}
    {E_0 + \frac{\delta+\sigma k^{2}}{2}} & {v.k}  \\
    {v.k} & {E_0 -\frac{\delta+\sigma k^{2}}{2}}
\end{pmatrix}
+\begin{pmatrix}
	i(\gamma+\gamma_i) & 0  \\
	0 & i\gamma_i
\end{pmatrix}
\end{equation}
The corresponding complex eigenvalues are given by:
 \begin{equation}\label{eq:neweigenvalues}
 E_{\pm}(k)=E_{0}+i\frac{\gamma}{2} + i\gamma_i \pm \sqrt{\left(\frac{\delta+\sigma k^{2}+i\gamma}{2}\right)^{2}+v^{2} k^{2}}
\end{equation}
In the main text, the FEM simulations (Figs.~1d,e) and the experimental data (Figs.~4a,b) of the real and imaginary parts are all fitted by \eqref{eq:neweigenvalues} with the same set of parameters:  $E_0=2.0$ eV, $\delta$ = 0.59 meV, $\gamma$ = 19.9 meV, $v=76.6$ meV.µm and $\sigma=-35$ meV.µm$^{-2}$. The only difference is that we use $\gamma_i=0$ to fit FEM results of ideal structure since there is no inhomogeneous broadening. However for experimental data, a non-zero value of  $\gamma_i=4$ meV is necessary to reproduce the observed broadening offset. The comparison between the two fittings, as well as the FEM and experimental results, are reported together in Fig.~\ref{figS5}.

